\documentclass[aps,prb,twocolumn,superscriptaddress]{revtex4-1}
\usepackage[dvips]{graphicx}
\usepackage{bm}        
\usepackage{amssymb,latexsym,amsmath,epstopdf,graphics,color,natbib}

\def\e{\mathop{\rm \mbox{{\Large e}}}\nolimits}

\begin{document}
\title[]{Transmittance of a subwavelength aperture  flanked by a finite groove array \\ placed near the focus of a conventional lens}
\author{F. Villate-Gu\'io}
\affiliation{Instituto de Ciencia de Materiales de Arag\'on and Departamento de F\'isica de la Materia Condensada, CSIC-Universidad de Zaragoza, E-50009 Zaragoza, Spain}
\author{F. de Le\'on-P\'erez}
\email[Email: ]{fdlp@unizar.es}
\affiliation{Instituto de Ciencia de Materiales de Arag\'on and Departamento de F\'isica de la Materia Condensada, CSIC-Universidad de Zaragoza, E-50009 Zaragoza, Spain}
\affiliation{Centro Universitario de la Defensa de Zaragoza, Ctra. de Huesca s/n, E-50090 Zaragoza, Spain}
\author{L. Mart\'in-Moreno}
\email[Email: ]{lmm@unizar.es}
\affiliation{Instituto de Ciencia de Materiales de Arag\'on and Departamento de F\'isica de la Materia Condensada, CSIC-Universidad de Zaragoza, E-50009 Zaragoza, Spain}
\begin{abstract}
One-dimensional light harvesting structures illuminated by a conventional lens are studied in this paper. Our theoretical study shows that high transmission efficiencies are obtained when the structure is placed near the focal plane of the lens. The considered structure is a finite slit-groove array (SGA) with
a given number of grooves that are symmetrically distributed with respect to
a central slit. The SGA is nano-patterned on an opaque metallic film. It is found that a total transmittance of 80 \% is achieved even for a single slit when (i) Fabry-Perot like modes are excited inside the slit and (ii) the effective cross section of the aperture becomes of the order of the full width at half maximum of the incident beam. A further enhancement of 8 \% is produced by the groove array. The optimal geometry for the groove array consists of a moderate number of grooves ($ \geq 4$) at either side of the slit, separated by a distance of half the incident wavelength $\lambda$. Grooves should be deeper (with depth $\geq \lambda/4$) than those typically reported for plane wave illumination in order to increase their individual scattering cross section. 
\end{abstract}


\maketitle 
 
\section{Introduction}
\begin{figure}
 \includegraphics[width=8cm]{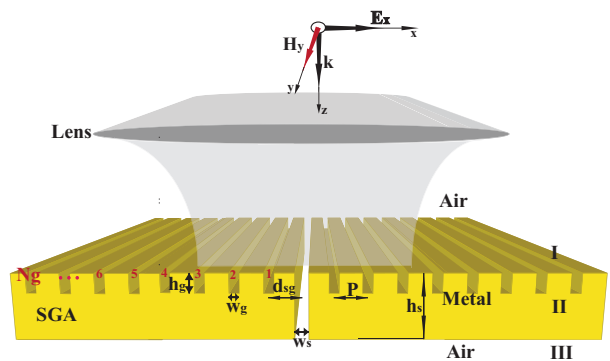}       
 \caption{(Color online). Schematic representation of a slit-groove array on a free-standing gold film (thickness $h_s$) illuminated by a thin cylindrical lens with focal distance $f$. A central slit of width $w_s$ is surrounded by $2N_g$ grooves (width $w_g$, depth $h_g$, period $P$), which are symmetrically distributed at either side of the slit. The distance from the slit to the first groove is called $d_{sg}$. The aperture is centered at the principal axis of the lens. 
}
\label{fig:scheme}
\end{figure}

Nanostructured metal films have important applications in sensing and photodetection due to their ability to collect and manipulate light at the nanoscale \cite{ThioNT02,ditlbacher02,OzbayS06,FJRMP10}. For instance, the high local density of states achieved inside a single hole can boost the fluorescence of molecules localized in the reduced volume of the hole \cite{RigneaultPRL05,WengerOE08}. A single aperture drilled on an opaque metal film can be surrounded by periodic structures, like in the slit-groove array (SGA) shown in Fig. \ref{fig:scheme}, 
to efficiently harvest light and subsequently
squeeze it through the aperture. 
The surface corrugation acts like an antenna that couples the incident light into surface modes, which are responsible for squeezing
the EM energy into the aperture \cite{FJRMP10}. The light harvesting process can be made more efficient by increasing the size of the system, provided that it does not exceed the propagation length of the surface modes. As a consequence of the light harvesting process, the power radiated to the farfield  can be one or two orders of magnitude larger than the power impinging on the aperture \cite{ThioOL01,LezecS02,HibbinsAPL02,FJPRL03,AkarcaAPL04,ThomasSSC04,JanssenPRL07,fvill01}.  The metal can also
be sculpted on the exit surface to
modify the re-emission pattern emerging from the aperture,
leading to the beaming phenomenon \cite{LezecS02,LMMPRL03}.  Phase compensation mechanisms produced by the outer corrugation allow the design of novel kind of lenses \cite{FangS05,LiuS07,SmolyaninovS07,MaAPL10,LuNC12}. 
The quasi-two dimensional nature of all such structures paves the way to its integration into optoelectronic devices with levels of miniaturization and operational speeds that could be never achieved with standard electronics \cite{CollinAPL03,IshiJJAPL05,YuAPL06,LauxNP08,DunbarAPL09,RossNP12}. 

A key feature of the optical response of a {\em subwavelength}  single slit is that its effective cross section ($\sigma$) is larger than the aperture size ($w_s$) \cite{FJRMP10}, which means that the aperture transmits more light than the one impinging directly on it. Even an {\em infinitesimal} narrow slit at resonance in a perfect electric conductor (PEC) film has a large $\sigma_{pw}=\lambda/\pi$ \cite{HarringtonAP80}, which is of the order of the wavelength $\lambda$ of the incident radiation. 

It is usually assumed in theoretical calculations that the system is illuminated by a plane wave (PW), in accordance with the broad wave front used in typical experimental conditions. In this paper we analyze the case that the PW front is focused with help of a conventional lens. Thanks to its macroscopic dimensions, the lens  collects an amount of energy several orders of magnitude larger than the one impinging directly on the aperture and focuses such energy in a spot with size of the order of the wavelength. When the size of the focus and the effective cross section of the aperture are of the same order, a large total transmittance ($T$) is expected for the new illumination conditions. However, to the best of our knowledge, this way of illuminating the system has not been examined previously. 

The first goal of this paper is to show the enhancement of $T$  when a subwavelength aperture is located at the focal plane of a lens. Furthermore, the density of EM energy near a subwavelength aperture provided by  the lens is so high that  one should wonder if a further enhancement of $T$ is still possible. Therefore, the second goal of this paper is to show that a groove array can increase the total amount of light transmitted through the aperture, even in the case that only a small fraction of the groove array is illuminated by the  central spot of the lens. Moreover, the sensitivity to the position of the SGA with respect to the focus of the lens is studied. We also provide ideal geometries for achieving high transmission intensities by optimizing the system shown in Fig. \ref{fig:scheme} under the new illumination conditions. 

The paper is organized as follow. Section \ref{MT} describes the theoretical framework. The optical response of a single slit is first considered in section \ref{sec:SS}. Optimal SGAs are discussed in section \ref{sec:SGA}. At the end of the paper the main conclusions are presented.

\section{Theoretical framework}
\label{MT}
Figure \ref{fig:scheme} describes the geometry of the system under study: a SGA illuminated by a cylindrical lens with focal distance $f$.
We assume that a $p$-polarized plane wave of wavelength $\lambda$ and electric field $\mathbf{E}_0$, parallel to the $x$ axis, impinges perpendicularly on a thin cylindrical lens fixed at $z=0$, see Fig. \ref{fig:scheme}. The fields of the lens are computed adapting the general method presented in Ref. \cite{Novotny} to the case of a cylindrical lens. This method consists in two main steps: (i) the fields near the optical lens  are formulated by the rules of geometrical optics within the paraxial approximation \cite{Goodman} and (ii) the lens fields are rigorously propagated to an arbitrary plane $z>0$ at the far-field by the means of the angular spectrum representation \cite{Novotny}. In this way, the non-vanishing components of the EM fields parallel to the metal plane can be written as 
\begin{equation}
\begin{aligned}
 E_x(x,z) &=& \frac{1}{\sqrt{4 \pi k_\lambda C_0}} \int_{-k_\lambda}^{k_\lambda} \e^{i \phi(k;x,z) } dk,  \\
H_y(x,z) &=&  \frac{1}{\sqrt{4 \pi k_\lambda C_0}} \int_{-k_\lambda}^{k_\lambda} Y_k \e^{i \phi(k;x,z) } dk, 
\end{aligned}
\end{equation}
where $\phi(k;x,z) \equiv k x+k_\lambda z+(f-z)k^2/2 k_\lambda$, $k_\lambda=2\pi/\lambda$ is the wave number of the incident wave, $Y_k \equiv k_\lambda/k_z \approx 2 k^2_\lambda/(2k^2_\lambda-k^2)$ is the  admittance of the propagating fields, and $C_0=\sqrt{2} \mbox{ArcSinh} (1) \approx 5/4$. The perpendicular component of the electric field, $E_z$, can be derived from Maxwell's equations. As we are computing the fields in the far-field of the lens, the above integrals are limited to propagating states $|k| \leq k_\lambda$.  The fields are normalized to render a unity total power behind the lens, 
\begin{equation}
P^{inc}_t =\int^{\infty}_{-\infty} \mbox{Re}[E_x \times H^*_y] \, dx= 1. \nonumber 
\end{equation}

\begin{figure}
   \includegraphics[width=9cm]{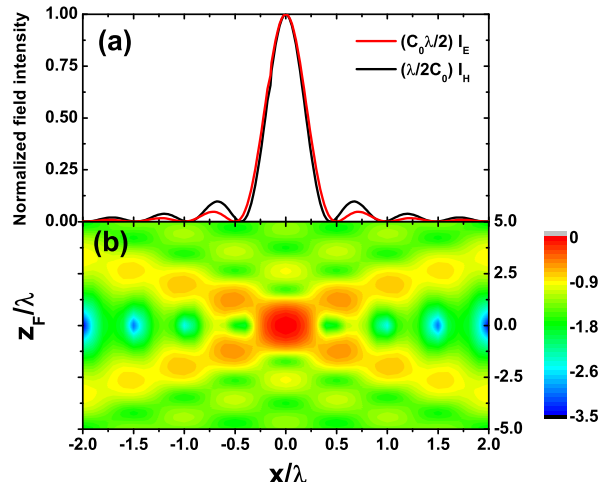} 
    \caption{(Color online). (a) Normalized intensities of the parallel components of electric ($C_0 \lambda I_E/2$) and magnetic ($\lambda I_H/2 C_0$) fields  at the focal plane of the lens. (b) Contour plot of $\lambda I_H(x,z_F)/2 C_0$ (in a logarithmic scale). Both $z_F$ and $x$ are in units of $\lambda$. 
}
\label{fig:fieldfoc}
\end{figure}

It is straightforward to compute the intensity of the $x$-component of the electric field $I_E(x,z)$ at the focal plane ($z=f$), 
\begin{eqnarray}
I_E(x,f)=|E_x(x,f)|^2 = \frac{2}{\lambda C_0} \mbox{sinc}^2 (k_\lambda x), \nonumber
\end{eqnarray}
where the function $\mbox{sinc}(x) \equiv \sin x/x$. This quantity has an absolute maximum equal to $2/(\lambda C_0)$ at $x=0$, secondary maxima at $x_n=(2n+1) \lambda/4$, and vanishes at the points $x_n=n \lambda/2$, where $n$ is an integer different from zero. The width of the main peak, defined as the distance between the two first minima, is equal to $\lambda$. The normalized intensity at the focal plane, $C_0 \lambda I_E(x,f)/2$,  is represented in Fig. \ref{fig:fieldfoc}(a). 

The intensity of $H_y$ at the focus is $I_H(0,f)=2C_0/ \lambda$. It is 56 \% larger than $I_E(0,f)$. $I_H(x,f)$ at other points of the focal plane has a more cumbersome expression that must be evaluated numerically. In contrast with PW illumination, $E_x$ and $H_y$ are no longer proportional after traversing the lens, although their normalized intensities show similar features, see Fig. \ref{fig:fieldfoc}(a). It is worth to notice, however, that the minima of $H_y$ are slightly shifted to  $x \lesssim n \lambda/2$.

Fig. \ref{fig:fieldfoc}(b) shows the intensity of the $y$-component of the magnetic field (in a logarithmic scale) in a region near the focus. The position relative to the focal plane is labeled as $z_F \equiv z-f$. Out of focus, the main peak is broader and has an intensity lower than at the focal point, although  the minima at $x_n \lesssim n \lambda/2$ show now a non-zero intensity. 

The optical response of the system is computed in the framework of the coupled mode method (CMM) \cite{FJRMP10}, which is briefly reviewed here for the sake of completeness. The CMM is based on a convenient representation of the EM fields in each region (waveguide modes and plane waves are used inside the defects and in the space surrounding the metal film, respectively). 

The metal is treated as a PEC. This approximation captures the main trends of the response of a metal at optical frequencies, is a good approximation in the infrared part of the spectrum, and provides practically exact results for the THz regime \cite{FJRMP10,fvill01}. A generalization of the results for the case of a real metal is straightforward \cite{FdLPNJP08} and does not change the main conclusions at which we arrive in the paper.
 
For a subwavelength aperture, convergence is rapidly achieved with a small number of waveguide modes, with the fundamental-mode approximation already showing an excellent agreement  with fully converged results \cite{FLTNJP08}. Therefore, only the fundamental propagating eigenmode of a planar waveguide is considered for slit and grooves, i.e. the electric field inside the indentation $\alpha$ is a linear combination of $\phi_\alpha \mbox{exp}(\pm i k_\lambda z)$, with $\phi_\alpha=w^{-1/2}_s$ and $w^{-1/2}_g$ for the slit and the $N_g$ identical grooves, respectively.

After matching the fields at the interfaces, 
the matching equations can be expressed as a function of the set $\{E_\alpha,E'_0\}$, which gives the amplitude of the waveguide modes right at the indentation openings: $ E'_0$ at the output side of the slit and $E_\alpha$ at the input side of slit ($\alpha=0$) and grooves ($\alpha=\pm1$, $\pm2$, $\ldots$, $N_g$). After some straightforward algebra, we end up with the following set of $2N_g+2$ equations for the unknowns  $\{E_\alpha,E'_0\}$ \cite{FJPRL03}: 
\begin{equation} \label{eq:SE}
\begin{split}
         (G_{\alpha \alpha}-\epsilon_\alpha)E_\alpha+\sum_{\beta \not = \alpha} G_{\alpha \beta} E_\beta -G_{\nu}E'_0 \delta_{\alpha 0} &=I_\alpha,  \\
           (G_{00}-\epsilon_0)E'_0 -G_{\nu}E_0 &=0.
\end{split}
\end{equation}
The function $\epsilon_\alpha=-i\cot(k_\lambda h_\alpha)$ takes into account the bouncing back and forth of the electromagnetic fields inside the defects. $G_\nu=-i\sin^{-1}(k_\lambda h_s)$ represents the coupling between the fields at the two sides of the slit. $G_{\alpha \beta}=\langle\varphi_\alpha| G_\sigma |\varphi_\beta\rangle$ is the projection onto wavefields $\varphi_\alpha$ and $\varphi_\beta$ of the green function $G(x,x')=\pi \lambda^{-1} H_{0}^{(1)}(k_\lambda |x-x'|)$, $H_{0}^{(1)}(x)$ being the Hankel function of the first kind \cite{FJRMP10}. The term $G_{\alpha \beta} E_\beta$ in Eq. (\ref{eq:SE}) can be viewed as the reillumination of indentation $\alpha$ by the EM field coming from indentation $\beta$, while $G_{\alpha \alpha}E_\alpha$ is a self-interaction term.

Only the independent term $I_\alpha$ is a function of the field provided by the lens. It gives the direct illumination over defect $\alpha$, centered at the position $x_\alpha$.  $I_\alpha$  is proportional to the overlap of the waveguide mode $\alpha$ and the incident EM field, and reads
\begin{equation}
\label{eq:Ia}
I_\alpha(x_\alpha,z)= \sqrt{\frac{w_s}{\pi k_\lambda C_0}} \int_{-k_\lambda}^{k_\lambda} Y_k \e^{i \phi(k;x_\alpha,z) } \mbox{sinc} (\phi_\alpha) \, dk, \nonumber
\end{equation}
where $\phi_\alpha=k w_\alpha/2$. 
For a very narrow aperture ($\mbox{sinc} (\phi_\alpha) \approx 1$) located at the focal plane, the illumination is proportional to the magnetic field, $I_\alpha(x_\alpha,f) \simeq 2 w_\alpha^{1/2} H_y(x_\alpha,f)$. At the focus, it simplifies to 
\begin{eqnarray}
\label{eq:I0f}
 |I_\alpha(0,f)| = 2 \left(\frac{2 w_\alpha C_0}{\lambda} \right)^{1/2} \approx \left(\frac{10 w_\alpha}{\lambda} \right)^{1/2}.
\end{eqnarray}  

Finally, the total transmittance $T$ is defined as the ratio of the energy power radiated to the far-field $P_{rad}$ and the total incident power $P^{inc}_t$. $T$ can be cast in a simple analytical form \cite{FJRMP10}
\begin{eqnarray}
\label{eq:T}
 T \equiv \frac{P_{rad}}{P^{inc}_t}=Re \bigl[G_\nu E_0 E'^*_0\bigr],
\end{eqnarray}
 where $E'^*_0$ is the complex conjugate of $E'_0$. The effective cross section of a slit of width $w_s$ is defined as 
\begin{eqnarray}
\label{eq:sigmaL}
 \sigma \equiv w_s \frac{P_{rad}}{P^{inc}_{slit}}= w_s  \frac{P^{inc}_t}{P^{inc}_{slit}}T,
\end{eqnarray}
where $P^{inc}_{slit}$ is power incident directly on the slit. For the particular case of infinitesimal slit at the focus, it is straightforward to compute the incident power,  $P^{inc}_{slit}/P^{inc}_t=2 w_s/\lambda$;	 therefore
\begin{eqnarray}
\label{eq:sigmaL0}
 \sigma=\frac{T}{2} \lambda.
\end{eqnarray}

\section{Single Slit}
\label{sec:SS}

In this section we study the optical response of a single slit of width $w_s$ drilled in a PEC film with thickness $h_s$. The spectral locations of its characteristic transmission peaks are associated with the resonant  condition 
\begin{eqnarray}
\label{eq:res}
 |E_0|=|E'_0|,
\end{eqnarray}
see Ref. \cite{FJRMP10}.  It gives rise to Fabry-Perot (FP) like modes, which are independent of the way the system is illuminated. We consider first the extreme subwavelength regime for the slit, $w_s \ll \lambda$. In this limiting case, FP modes occur when the metal thickness is an integer multiple of half the wavelength, $h_s=n \lambda/2$ \cite{FJPRL03}. As mentioned above, an infinitesimally narrow slit has a large ($\sim \lambda$) effective cross section $\sigma_{pw}=\lambda/\pi$ when it is at resonance. This expression was first reported in Ref. \cite{HarringtonAP80} and later re-derived in the framework of the CMM \cite{FJRMP10}  by taking the limit $w_s \ll \lambda$ in Eq. (\ref{eq:SE}). 

\begin{figure}
\includegraphics[width=9cm]{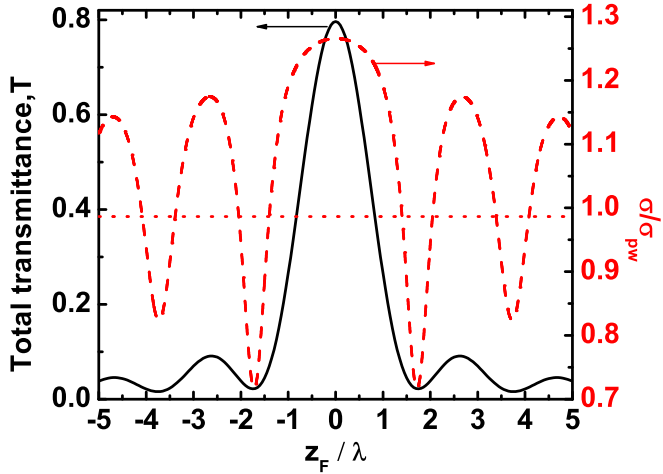}
\caption{(Color online). Total transmittance $T$ (black-solid line) and effective cross-section $\sigma$ (red-dashed line, normalized by $\sigma_{pw}=\lambda/\pi$) versus $z_F/\lambda$ for a slit of  width $w_s=0.1 \lambda$ in a metal film of thickness $h_s=0.37\lambda$ illuminated by a lens. The red-dotted horizontal line represents $\sigma_{pw}(w_s=0.1 \lambda)=0.99 \sigma_{pw}$ for the same slit  under PW illumination.}
\label{fig:SS}
\end{figure}

PW illumination has been assumed in those previous calculations. However, when the incident light is focused by the lens on the infinitesimal aperture, a closed-form expression can be derived for the total transmittance. 
We start from the general expression for the total transmittance at  resonance
\begin{eqnarray}
 T_{res}=\frac{\lambda |I_0|^2}{4\pi w_s}, \nonumber
\end{eqnarray} 
derived in Ref. \cite{FJRMP10} for the extreme subwavelength regime. It was obtained using  the resonant condition (\ref{eq:res}) in Eq. (\ref{eq:T}) and taking the limit $w_s \ll \lambda$ in the expression for $G_{00}$. Assuming that the infinitesimal slit is at the focal point of the lens ($x_0=z_F=0$) and using Eq. (\ref{eq:I0f}) for $|I_0|$, we find that 
\begin{eqnarray}
 T_{res}=\frac{2 C_0}{\pi} \approx \frac{5}{2\pi} \approx 0.8. \nonumber
\end{eqnarray}
Such large value of $T_{res}$, which is independent of both $w_s$ and $\lambda$ for a very narrow aperture, is the main result of this paper. We recall that, in contrast with this 80\% of efficiency,  the total transmittance for a PW impinging directly on metal surface is vanishing small due to the infinite extension of its wave front. In practice, the PW transmittance should be normalized to the power incident on a finite area, like the area of the aperture \cite{FJRMP10}. 

The enhancement of $T$ can be explained by the large effective cross section of the slit, which is easily obtained after replacing $T_{res}$ in Eq. (\ref{eq:sigmaL0}),
\begin{eqnarray}
\label{eq:sigmaS}
 \sigma=C_0 \sigma_{pw} \approx \frac{5}{4} \sigma_{pw}\approx 0.40 \lambda.
\end{eqnarray} 
It approaches the full width at half maximum  (FWHM) of the intensity of the incident field, which is equal to FWHM$\approx 0.44 \lambda$ in Fig. \ref{fig:fieldfoc}(a).
Such small increment of 25 \% with respect to $\sigma_{pw}$ is enough to allow that a 80 \% of the focused light can be transmitted through the aperture.

All previous analytical results have been derived for a very narrow subwavelength aperture with $w_s \ll \lambda$. We study now a {\em finite} subwavelength slit with a typical experimental width, $w_s=0.1 \lambda$.  For $w_s>0$, FP-like modes at fixed $\lambda$ are excited in thinner metal films, $h_s < n\lambda/2$ \cite{TakakuraPRL01}. So we get a resonant thickness of $h_s=0.37\lambda$ for $w_s=0.1 \lambda$ and $n=1$.

  Fig. \ref{fig:SS} shows both $T$ and $\sigma/\sigma_{pw}$ as a function of $z_F$ for such slit. At the focal point, we have that $T \approx 0.8$ and $\sigma \approx 1.27 \sigma_{pw}$. Such values are slightly larger than (but in excellent agreement with) those obtained for an infinitesimal aperture.  The difference is on the second decimal place. Notice that for the finite slit under PW illumination $\sigma_{pw}(w_s>0)$ is slightly smaller than for an infinitesimally narrow slit. We find $\sigma_{pw}(w_s=0.1 \lambda)=0.99 \sigma_{pw}$. This value is represented with a red-dotted horizontal line in Fig. \ref{fig:SS}.

As a function of $z_F$, $T$ shows a main peak at the focal plane of the lens ($z_F=0$) and secondary peaks of lower intensity when the slit  is out of focus. This behavior can be explained by the non-monotonous dependence of the magnetic field incident on the slit. Extreme values of both $|H_y|^2$ and $T$ occur at the same $z_F$, c.f. Figs. \ref{fig:fieldfoc}(b) and \ref{fig:SS}(a). The oscillation of $\sigma$  around the constant value for PW illumination can be explained in a similar way.

\begin{figure}
\includegraphics[width=9cm]{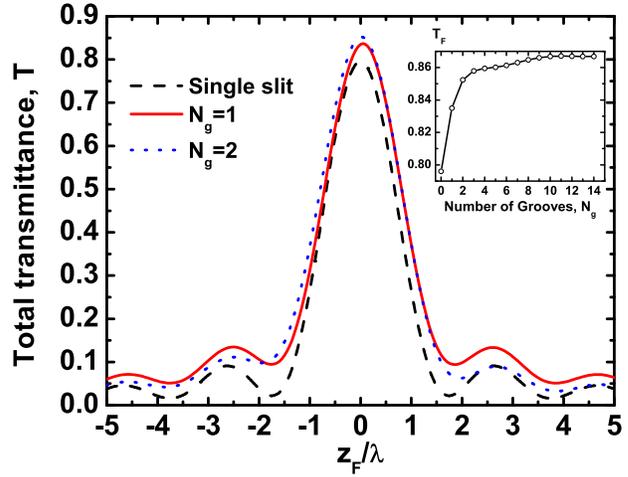}
\caption{(Color online). (a) $T$ versus $z_F/\lambda$ for a SGA with $N_g=0,1,2$ grooves optimized at the focus of the lens. The optimal geometry of the grooves is $w_g=0.1 \lambda$, $h_g=0.3 \lambda$, $d_{sg}=0.4 \lambda$, and (for $N_g>1$) $P=0.46 \lambda$. The slit is the same than in Fig. \ref{fig:SS}. The inset shows the intensity of the main peak at the focus ($T_F$) as function of the number of grooves ($N_g$).}
\label{fig:io}
\end{figure}

\section{Slit-groove array}
\label{sec:SGA}
We study now the total transmittance of a SGA near the focal plane of the lens. We assume that all  grooves are identical, with depth $h_g$, width $w_g$, period $P$, and distance $d_{sg}$ from the first groove to the central slit. As in Fig. \ref{fig:SS}, we consider a metal film with thickness $h_s=0.37\lambda$ and a slit of  width $w_s=0.1 \lambda$ centered at the principal axis of the lens.

Fig. \ref{fig:io} shows $T$ as a function of $z_F$, near the focal region, for a SGA with $N_g=0$ (black-dashed line), 1 (red-solid line), and 2 (blue-dotted line) grooves at either side of the central slit. The inset shows the transmittance at the focus ($T_F$) as a function of the number of grooves $N_g$. 

The response of the system has been optimized for providing a high transmission intensity when it is placed at $z_F=0$. In the optimization process, we can follow design rules similar to those developed in Ref. \cite{fvill01} for PW illumination. Thus, the largest transmittance is obtained when the Fabry-Perot mode of the slit (discussed in the previous section)  is located at the same spectral position than the mode excited in the groove array \cite{FJPRL03}. The  mode in the array is a function of the groove geometry, its  optimal periodicity guarantees that the light re-emitted by the grooves reaches in phase the other grooves and the central slit. The interaction between the slit and the groove array can be further controlled  by modifying $d_{sg}$. 

The main difference we find with respect to the optimal response for PW illumination is that groove cavity modes are not excited when the grooves are illuminated by the lens. Cavity modes for a groove with an infinitesimal width ($w_g \ll \lambda$) are obtained for a groove depth equal to $h_g=n\lambda/4$, where $n$ is an integer \cite{FJPRL03}. For a finite width, the first order groove cavity mode appears for $h_g<\lambda/4$ \cite{fvill01}; when $w_g$ tends to zero, the ideal $h_g$ approaches $\lambda/4$ from below. However, in the presence of the lens, the first order mode for a finite width is found for $h_g>\lambda/4$, so the groove cavity mode is not excited. Moreover, when $w_g$ tends to zero, the ideal $h_g$ approaches $\lambda/4$ from above. Grooves need to be deeper than in the case of PW illumination in order to increase their scattering cross section. In this way, the groove compensate the reduction of the intensity of the incident light out of focus, see Fig. \ref{fig:fieldfoc}(a). On the other hand, we find that, for an optimal reillumination of the central slit, the distance between neighboring grooves should be of the order of $\lambda/2$ (see the discussion below). A set of geometrical parameters that satisfies all such requirements  are given in the caption of Fig. \ref{fig:io}. 

We observe in Fig. \ref{fig:io} that the intensity $T_F$ of the main peak for the SGA is larger than for a single slit. Moreover, $T_F$ increases with the number of grooves. An enhancement of 8 \% with respect to the single slit is found. However, $T_F$ is saturated for a moderate number of grooves; practically not further enhancement is found for $N_g>4$ grooves. The intensity of the secondary peaks decreases with $N_g$ for $N_g>1$. However, the main peak shows  an intensity  $T$ that is still equal to $T_F/2$ at $z_F=0.9 \lambda$. Its FWHM=$1.8 \lambda$ is practically independent of $N_g$ when $N_g>4$ grooves.


\begin{figure}
\includegraphics[width=9cm]{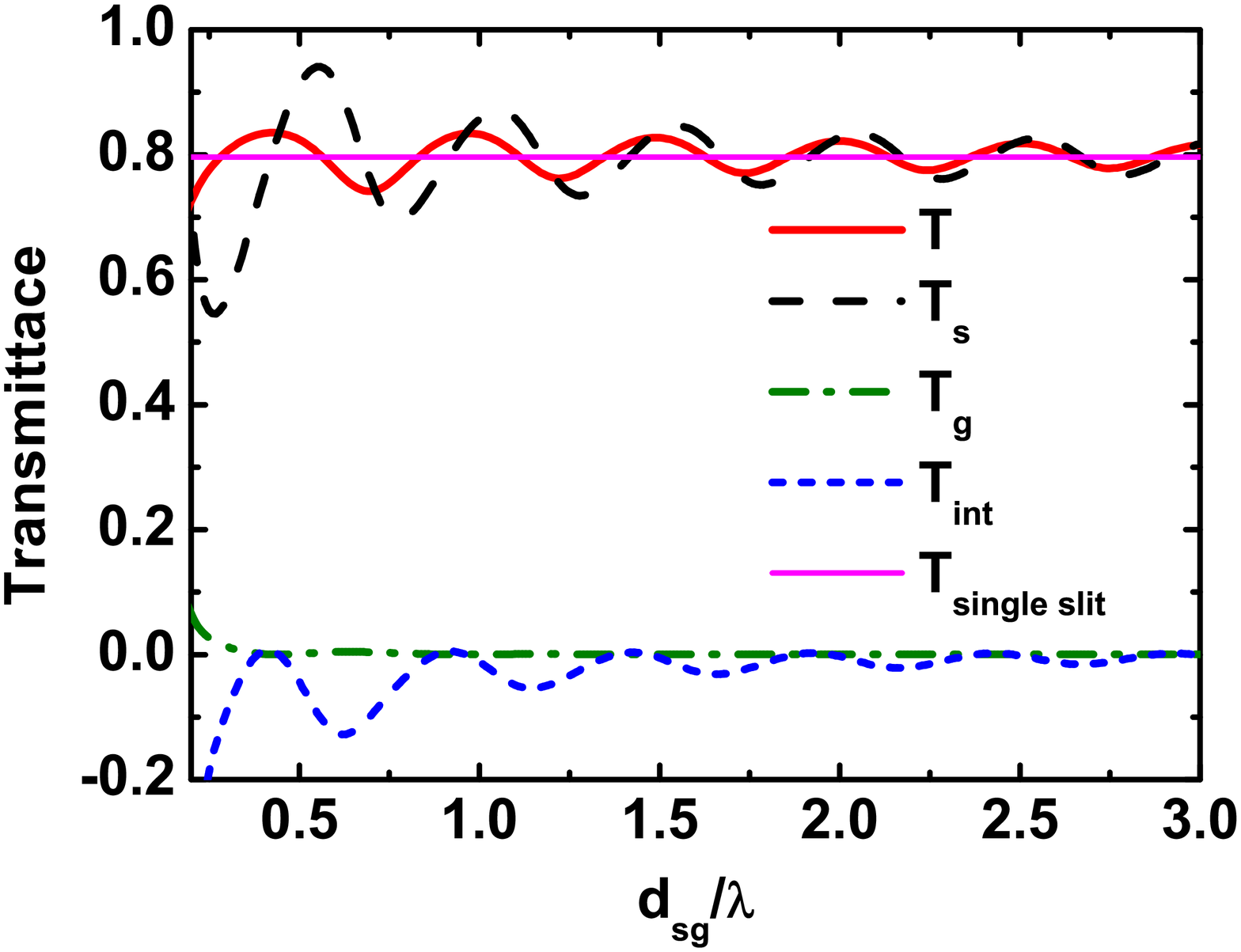}
\caption{(Color online) $T$, as well as its constituents terms from Eq. (\ref{eq:TNg1}), as a function on the slit-groove distance, $d_{sg}$, for a SGA with $N_g=1$ at focal plane of the lens and the same geometrical parameters of Fig. \ref{fig:io}. The magenta horizontal line represents $T$ for the single slit.}
\label{fig:inter}
\end{figure}

 In order to gain physical inside, we make a careful ana\-ly\-sis of the transmittance of the SGA with  $N_g=1$. For this simple case (only two identical grooves are interacting with the central slit),  closed-form expressions for the amplitudes of electric field at the input and output sides of the slit, $E_0$ and $E'_0$, respectively, can be easily found by solving the system of Eqs. (\ref{eq:SE}). Replacing both $E_0$ and $E'_0$ in Eq. \ref{eq:T}, the total transmittance can be written as
\begin{eqnarray}
\label{eq:TNg1}
 T=T_s+T_g+T_{int}, 
\end{eqnarray}
 where $T_s$ is the transmittance when only the central slit is illuminated ($I_1=0$), $T_g$ is obtained when only the grooves are illuminated ($I_0=0$), and $T_{int}$ is an interference term. 

Fig. \ref{fig:inter} shows $T$ at $z_F=0$, as well as its constituents terms, as a function of $d_{sg}$ for a SGA with $N_g=1$. We find that the slit receives most of the light directly from the lens, while only a small fraction is coming from the grooves ($T_s \gg T_g$). In fact, $T_g$ decays very fast with $d_{sg}$, being practically negligible for $d_{sg}>0.3 \lambda$. The oscillation of $T_s$ is related to the conditions of interference for the reillumination process. A similar behavior has been already reported in Refs.  \cite{HibbinsAPL02,SoenninchensenAPL00,SchoutenPRL05,LalannePRL05,AlarverdyanNP07,PacificiOE08,HafeleAPL12}. The oscillation of $T_{int}$ has a similar origin. $T_{int}$ is in antiphase with $T_s$, leading to a total transmittance $T$ that oscillates around the value obtained for a single slit. As a result of the interference between slit and grooves, we find that the ideal  positions of the grooves are those located near the minima of the incident field, $d_{sg} \approx n\lambda/2$, see Fig. \ref{fig:fieldfoc}. 

\begin{figure}
\includegraphics[width=9cm]{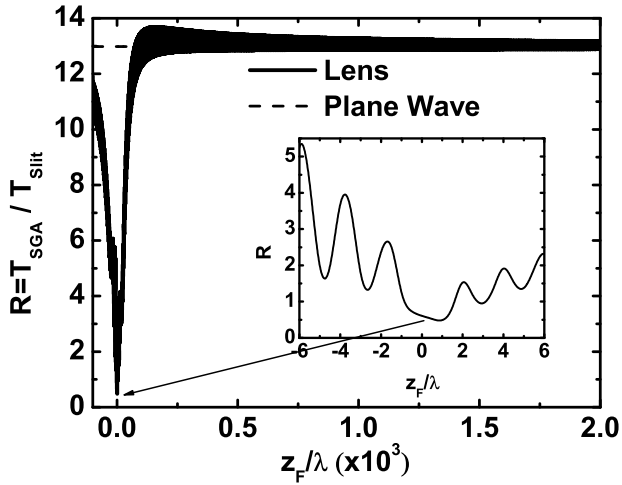}
\caption{Ratio $R=T_{SGA}/T_{slit}$ for a SGA optimized for PW illumination and illuminated either by a lens (solid line) or directly by a PW (dashed lined). The inset shows a zoom of $R$ near the focal plane of the lens. In the calculations, we use the same slit than in Fig. \ref{fig:SS} and 6 grooves at each side of the slit with optimal geometry: $w_g=0.1 \lambda$, $h_g=0.14 \lambda$, $P=0.95 \lambda$, and $d_{sg}=0.87 \lambda$. The total size of the SGA is $L=11.2 \lambda$.}
\label{fig:tsgss}
\end{figure} 

Finally, it is worth to stress that a SGA optimized for PW illumination renders a poor response at the focal plane of the lens. Figure \ref{fig:tsgss} illustrates this behavior for a SGA with 6 grooves at each side of the slit. Optimal geometric parameters are given in the figure caption. Grooves optimized for PW illumination have $d_{sg},P \sim \lambda$  and are shallower than those previously optimized at the focal plane of the lens, c.f. the geometries reported in Figs. \ref{fig:tsgss} and \ref{fig:io}. 

In order to characterize the response of the SGA when it is placed off focus, we define the ratio $R=T_{SGA}/T_{slit}$. It is equal to $R_{PW}=12.98$ when the SGA optimized for PW illumination is illuminated by a PW (in this case $T$ is normalized to the power incident on the size of the system). Figure \ref{fig:tsgss} shows $R$ for the same system but near the focus of the lens and compares this curve with the constant value $R_{PW}$. Notice that both the numerator and the denominator of $R$ vary now with $z_F$. 

We observe that $R<R_{PW}$ near the focal region, having a deep minimum at $z_F=\lambda$, see inset to Fig. \ref{fig:tsgss}. Looking back at the field impinging on the SGA, see Fig. \ref{fig:fieldfoc}, we find that when the SGA is near the focus most of the grooves are outside the hot spot of the lens, so the light collected by them is not enough to achieve the same efficiency than for a SGA illuminated with a PW. Moreover, $R<1$ at the focus, i.e. the SGA optimized to operate under PW illumination, when placed at the focus of a lens, is less efficient than a single slit under the same illumination. Oscillations in $R$ (see inset) are due to the non-uniform illumination provided by the lens. 

By moving the SGA out of focus, the incident light reaches more grooves and therefore $R$ starts to raise. $R$ reaches its maximum value when the whole SGA is illuminated by the lens. Notice, however, that at such large relative distance $T$ is rather small due to the reduction of the power directly incident on the slit.  Far away from the focus, the hot spot is so broad that it behaves like a plane wave and $R$ tends to $R_{PW}$.

\section{Conclusions}
We have studied the optical response of a slit-groove array illuminated by a conventional cylindrical lens. The lens enlarges the effective cross section of the central slit by focusing the incident radiation on its opening. An effective cross-section 25 \% larger than for plane-wave illumination together with the confinement of the incident field leads to a total transmittance at the focus $T_F=80$ \% for the single slit. An optimal groove array further enhances $T_F$ in 8 \%. The transmittance $T$ decreases with the distance relative to the focal plane $z_F$; however, at a distance to the focus $z_F=0.9 \lambda$, $T$ is still equal to $T_F/2$. Moreover, we have found that a slit-groove array optimized far away from the focus (i.e. for PW illumination) renders a low efficiency when it is at the focal plane of a lens.

The ideal geometry of the slit allows that the well known Fabry-Perot like modes are excited inside it.  Grooves should be deeper (with depth $\geq \lambda/4$) than those typically reported for PW illumination in order to increase their scattering cross section. A moderate number of grooves  ($\geq 4$) is needed for achieving high intensities. The ideal distance between grooves is  close to $\lambda$ for PW illumination. However, grooves should be separated by a distance of $\sim \lambda/2$ at the focal plane of the lens. 

We hope that the present work could motivate further experimental and theoretical studies of the interaction of a SGA with both conventional lenses and metalenses, as well as future applications in the design of optical and infrared detectors.

\begin{acknowledgments}
The authors gratefully acknowledge financial support by European Projects EC FP7-ICT PLAISIR Project  Ref. 247991 and the Spanish Ministry of Science and Innovation project MAT2011-28581-C02-02.
\end{acknowledgments}

%
\end{document}